# Microscopic origin of the spin-splitting in altermagnets

Suyoung Lee[1,*], Minjae Kim[2,*] and Changyoung Kim[1†]

[1]Department of Physics and Astronomy, Seoul National University, Seoul 08826, Korea
[2]School of Computational Sciences, Korea Institute for Advanced Studies, Seoul 02455, Korea

*SL and MK contributed equally to this work.
[†] Corresponding author: changyoung@snu.ac.kr

## Abstract:

Altermagnets, characterized by spin-split bands without net magnetization, have recently emerged as a promising platform for spintronics. However, their microscopic mechanisms remain elusive, often relying on abstract group theory. In this work, we present an intuitive and pedagogical framework to understand the origin of spin splitting in altermagnets. We identify two essential ingredients: (1) alternating spin-polarized wavefunction localization on sublattices, and (2) broken translational symmetry caused by distortions in non-magnetic ion cages. We discuss a minimal model Hamiltonian based on an atomic exchange-driven spin splitting and anisotropic hopping that captures these effects and reproduces the hallmark features of altermagnetic band structures, including nodal spin degeneracies and large spin splittings. Our model is further validated by ab initio calculations on $MnF_2$. By demystifying the microscopic origins of altermagnetism, our work bridges symmetry analysis and material realizations, shedding light on practical designs of altermagnetic spintronic devices.

# I. Introduction

The third class of magnetism, dubbed as altermagnetism [1], has been a recent hot topic in the field of magnetism. Altermagnets have characteristics of both ferro- (FM) and antiferro-magnetism (AFM): spin split bands (thus broken time reversal symmetry) and zero net magnetization, respectively. These traits are important not only in the fundamental scientific point of view but also for spintronic applications. The spin split bands with accidental and symmetry protected spin degenerate nodes can lead to topological features such as Weyl nodes which are actively discussed in recent studies [2–4]. As for spintronic applications, altermagnets are supposed to have merits of both FM and AFM (strong signal and high operating speed, respectively) [5]. The anisotropic spin polarized Fermi surfaces can generate unconventional spin splitter torque (SST) even in the absence of spin-orbit coupling. In addition, large anomalous Hall effect (AHE) allowed in some altermagnets provides an electrical readout of altermagnetic order [6,7].

While these novel properties make altermagnets an attractive topic, intuitive understanding of altermagnetism is still lacking in the community. Initial works on the mathematical definition of "altermagnetism" were based on spin group analysis [8,9]. Approaches to distinguish compensated magnets with broken time reversal symmetry have been also made in terms of magnetic symmetry [10–12]. Although such symmetry-based description of altermagnetism provides formal thoroughness, it makes the topic less accessible for those who are not familiar with the group theory. In addition, the symmetry group analysis does not show degree of the spin splitting mechanism (symmetry does not provide information on 'how much'). Approaches based on tight-binding model began to appear only recently [2,13]. Recent works have introduced magnetic multipole as the "order parameter" driving altermagnetism, which not only provides a more comprehensive understanding but also naturally accounts for macroscopic phenomena arising in this class [14,15]. However, they still fall short of providing a fully intuitive picture. An intuitive understanding on the other hand often allows us to see hidden physics and also have an appropriate direction for materials design. Therefore, a more pedagogical approach to altermagnetism may be desired.

The purpose of this work is to establish a pedagogical approach to the microscopic mechanism of the spin split bands in altermagnets. We show intuitively that the main mechanism involves two factors; an AFM wave function for a particular spin direction has significant weight at every other site, and the local structure of the non-magnetic ions has distortion in an alternating way. The result is that the wave function with a spin preferentially resides at sites with a particular distortion, leading to momentum dependent spin splitting. We set up a Hamiltonian based on the picture to show that the picture naturally explains the known electronic structures of altermagnets.

## II. Concept of altermagnetism

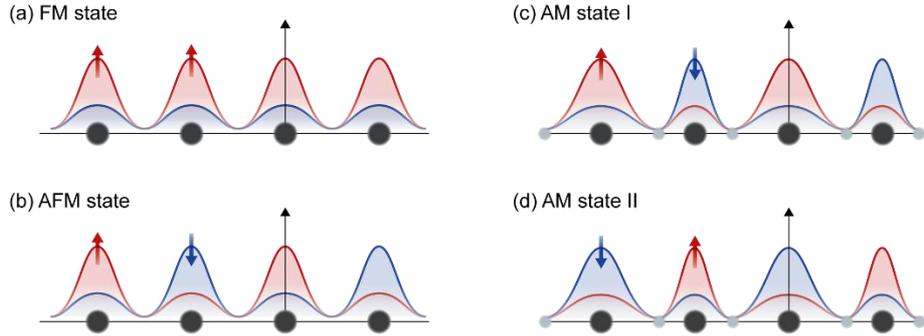

**Figure 1** Schematic spin resolved electronic wave functions in (a) ferromagnetism (FM), (b) antiferromagnetism (AFM), and two types of magnetic states (c) I and (d) II in altermagnetism (AM). Black and grey circles denote magnetic and nonmagnetic ions, respectively.

The first step towards intuitive understanding of the altermagnetism is to know the shape of the spin dependent wave function. We plot in Fig. 1 schematic spin-resolved electronic wave functions for three representative magnetic states: ferromagnetism (FM), antiferromagnetism (AFM), and altermagnetism (AM). In the case of ferromagnetism, shown in Fig. 1(a), the spatial distribution of spin dependent electronic wave function clearly shows spin polarization and thus breaking of time-reversal symmetry. An important aspect to note is that, for each spin, the electron density is evenly distributed over the atomic sites. Such is not the case for antiferromagnetism, illustrated in Fig. 1(b). The wave functions for two spin directions occupy every other site in an alternating way, that is, two spins occupy sublattices in a *mutually exclusive way*. As we will see, this is a key ingredient for the spin split bands in AM. In spite of such distribution, the environment for each magnetic sublattice is the same. This results in PT symmetry of AFM, where P represents the inversion operator between magnetic sublattices and T denotes the time-reversal operator. The PT symmetry leads to absence of spin-dependent splitting throughout the Brillouin zone.

In contrast, the local environment around magnetic ions changes alternatingly in AM due to the arrangement of nonmagnetic ions, as depicted in Fig. 1(c). That is, the local environment for up-spin dominant sublattice is different from the down-spin dominant sublattice due to the presence and arrangement of nonmagnetic ions (filled grey circles). Therefore, the energy of the up-spin state is different from that of the down-spin, leading to a spin-splitting. In more formal terms, the spin-split band structure appears due to two essential factors: (1) the presence of non-magnetic ions, which break translational symmetry between magnetic sublattices, and (2) the existence of opposite spin polarizations in the two magnetic sublattices. An important aspect is that the configuration allows two different AM states as illustrated in Figs. 1(c) and 1(d). Note that a spin occupies different magnetic sublattices in the two magnetic states. Therefore, different magnetic states exhibit opposite signs of spin splitting.

It is worth emphasizing that the above microscopic perspective on the origin of spin splitting in systems with collinear AFM order remains valid to an extent even when the defining symmetries are reduced or altered. Specifically, when a three-dimensional (3D) altermagnet is thinned down to two-dimensional (2D) limit, the system may no longer qualify for altermagnetism; for instance, a single-unit-cell-thick (110) film of $RuO_2$ with a collinear AFM order does not conform to altermagnet due to broken translational symmetry. Nonetheless, the microscopic approach outlined above still clearly predicts spin-split bands. This discrepancy highlights a promising direction for future investigation.

## III. Domains and domain boundary

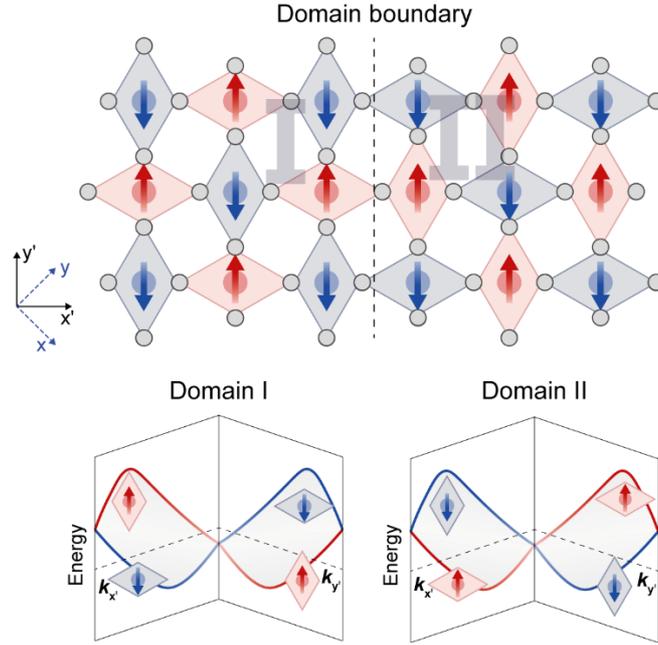

**Figure 2** Domains and domain boundaries in AM. The upper panel presents two domains and the domain boundary (dashed line). Two coordinate systems with axes labeled xy and x'y' are also defined. Band structures from the two domains are presented in the lower panel. Shown also in the lower panel is the magnetic sublattice rhombus unit associated with each band.

The aforementioned multiple AM states results in formation of AM domains. Figure 2 is a schematic illustration of formation of AM domains. For the two domains of I and II in the upper panel of Fig. 2, it is seen that the spin direction at the same sublattice is opposite; while up-spins are located in the rhombus elongated along $k_{x'}$ in domain I, they reside in the elongated one along $k_{y'}$ in domain II. This is a two-dimensional version of the configuration shown in Figs. 1(c) and 1(d). It also shows why we need two magnetic ions in a unit cell.

The lower panel shows the corresponding band structures from the two domains. For the band structure from domain I, spin-splitting orders are opposite for the $k_{x'}$ and $k_{y'}$ directions as is well known. Also shown in the figure is the shape of the magnetic sublattice units for a spin-split band. Along the $k_{x'}$ direction, up-spin electrons are located in the rhombus *elongated along $k_{x'}$* in the upper band. In contrast, down-spin electrons reside in the rhombus *elongated along $k_{y'}$* for the states along the $k_{y'}$ direction. On the right-hand side, configurations for domain II are illustrated. Comparing the two band structures, one can see that the band energy is determined by the momentum and elongation directions – whether they are parallel or perpendicular to each other. Therefore, it is the distortion of the sublattice unit that determines the energetics of the AM states.

Finally, we wish to briefly discuss the domain boundary. Existence of two domains inevitably leads to domain boundaries. The opposite spin polarization between the two domains generates dislocation of the band structure in the domain boundary. In equations (1) – (3) in the following section, we show that changing the sign of $h_{eff}$, the intra-atomic exchange which induces the local spin polarization, is equivalent to changing the sign of $\delta t$, the non-magnetic ion driven PT symmetry breaking term from hopping anisotropy. Based on this equivalence as well as the relation between $\delta t$ and the distortion of a magnetic sublattice unit, we predict that the lattice degree of freedom from non-magnetic ion's distortion must have strong dislocation at AM boundaries. This dislocation in altermagnetism is distinct from that of the magnetic domain boundaries in antiferromagnetism for which an inter-site magnetic exchange is the source of the dislocation, usually with a much smaller energy scale in comparison to the energy scale of $\delta t$ and $h_{eff}$ in AM.

## IV. Microscopic model Hamiltonian

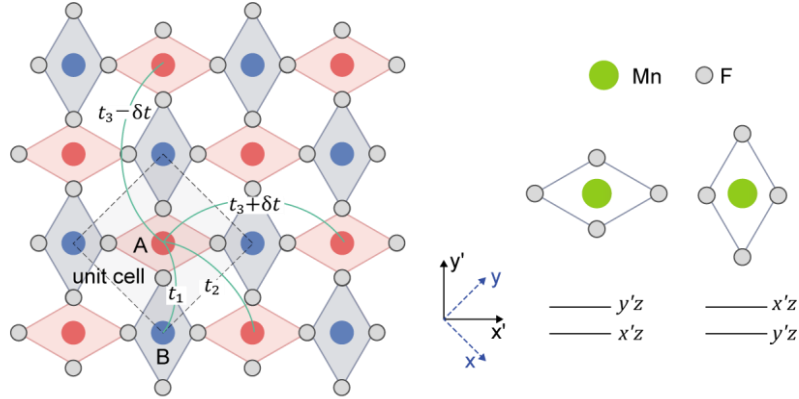

**Figure 3** Minimal model and orbital energy levels. (Left panel) Red and blue spheres are magnetic ions with opposite spin directions, with the basis composed of A and B. Rhombuses represent distorted squares of non-magnetic ions surrounding magnetic ions. $t_i$ is the electron hopping amplitude between $i$-th nearest neighboring magnetic ions. Due to the local distortion of non-magnetic ions, the third nearest hopping amplitude anisotropic, $t_3 \pm \delta t$. Note that this configuration represents a particular domain and one can obtain the other domain by flipping the spins. (Right panel) The distortion driven differentiations of $x'z/y'z$ orbitals in the two types of magnetic sublattices, due to the hybridization between magnetic ion (Mn) and non-magnetic ions (F) in an altermagnetic MnF$_2$. For the coordinate axes, refer to the definition in Fig. 2.

To address the microscopic mechanism of AM, we consider a minimal model on a two-dimensional square lattice from ref. [16], as depicted in Fig. 3. This minimal model encapsulates the essential characteristics of AM, including time-reversal symmetry breaking in the reciprocal momentum space, a zero net magnetic moment in the real space, the crucial role of non-magnetic ions, and the interplay of spin and lattice symmetries in generating spin splitting within the band structure.

The unit cell of this model comprises two types of magnetic ions, denoted as A and B, which exhibit opposite spin polarizations. Each magnetic ion is surrounded by rhombus-shaped arrangements of non-magnetic ions in a corner-sharing configuration. The model assumes a single orbital per magnetic ion, with the multi-orbital band structure recoverable by extending the Hilbert space. The hopping integral of an electron between the $i$-th nearest neighboring magnetic ions is denoted as $t_i$. Due to the elongation (shortening) of the non-magnetic ion's square along diagonal directions, the third-nearest-neighbor hopping is modified by an additional term $\delta t$. In the elongated (shortened) directions, electron delocalization (localization) results in modified hopping amplitudes of $t_3 + \delta t$ ($t_3 - \delta t$), respectively. This modification of the third-nearest-neighbor hopping term breaks the translational symmetry between the A and B sublattices.

Here, we wish to discuss in more detail why we have such anisotropic hopping amplitudes. For realistic materials, such as MnF$_2$ depicted in Fig. 3, the distortion of non-magnetic ions results in rotated point group symmetries for the two sublattices. The hybridization between the orbitals of magnetic and non-magnetic ions adheres to these point group symmetries, thereby disrupting the translational symmetry between the A and B sublattices. For instance, as illustrated in Fig. 3, the $x'z/y'z$ orbitals of Mn($d$) in MnF$_2$ exhibit oppositely modulated hopping between the A and B sublattices, leading to the emergence of a $\delta t$ term for each orbital. In particular, the $p-d$ hybridization induced electronic hopping, which is diagonal in spin, is stronger in a specific direction when (i) the $d$ orbital has a lobe in the direction for the stronger hybridization, and (ii) the $p-d$ bond length is elongated in the direction for the delocalization of the electrons in the $p-d$ hybridized orbital. Due to these factors, for the A (B) sublattice, the hopping along the $x'$ ($y'$) direction is enhanced compared to the $y'$ ($x'$) direction for the $x'z$ ($y'z$) orbitals.

With the hopping amplitudes as defined in Fig. 3, we can discuss our minimal model Hamiltonian presented in Eqn (1). The basis of $[A\uparrow, B\uparrow, A\downarrow, B\downarrow]^T$ constructs the Hilbert space. The reciprocal momentum-dependent coefficients are given in Eqn (2) following the discussion in Fig.3, where $\mu$ is the chemical potential of the system. In addition, $h_{eff}$ is an atomic exchange-driven spin splitting term from the emergence of the antiparallel spin polarization in magnetic ions of A and B. This atomic exchange-driven field has an order of few eV for *d*-orbitals of transition metal elements. The matrices of $I$, $\tau_x$, $\tau_z$, $\tau_z \otimes \sigma_z$, are also given in Eqn (2).

$$H = \epsilon_k \cdot I + t_{k,x} \cdot \tau_x + t_{k,z} \cdot \tau_z + h_{eff} \cdot (\tau_z \otimes \sigma_z) \tag{1}$$

where $\epsilon_k = 2t_2(\cos(k_x) + \cos(k_y)) + 4t_3(\cos(k_x)\cos(k_y)) - \mu$,

$t_{k,x} = 4t_1\left(\cos\left(\frac{k_x}{2}\right)\cos\left(\frac{k_y}{2}\right)\right)$, $t_{k,z} = -4\delta t(\sin(k_x)\sin(k_y))$,

$$I = \begin{bmatrix} 1 & 0 & 0 & 0 \\ 0 & 1 & 0 & 0 \\ 0 & 0 & 1 & 0 \\ 0 & 0 & 0 & 1 \end{bmatrix}, \tau_x = \begin{bmatrix} 0 & 1 & 0 & 0 \\ 1 & 0 & 0 & 0 \\ 0 & 0 & 0 & 1 \\ 0 & 0 & 1 & 0 \end{bmatrix}, \tau_z = \begin{bmatrix} 1 & 0 & 0 & 0 \\ 0 & -1 & 0 & 0 \\ 0 & 0 & 1 & 0 \\ 0 & 0 & 0 & -1 \end{bmatrix},$$

and $\tau_z \otimes \sigma_z = \begin{bmatrix} 1 & 0 & 0 & 0 \\ 0 & -1 & 0 & 0 \\ 0 & 0 & -1 & 0 \\ 0 & 0 & 0 & 1 \end{bmatrix}$ \tag{2}

Then, the band structure of the minimal model, obtained by solving Eq. (1), becomes

$$\epsilon_{\alpha=\pm,\sigma} = \epsilon_k \pm \sqrt{(t_{k,x})^2 + (t_{k,z} + h_{eff}\sigma)^2} \ . \tag{3}$$

The indices $\alpha$ and $\sigma$ denote the sign of the eigenvalue and the spin, respectively. The spin-splitting term in the momentum space from the equation is related to $2t_{k,z}h_{eff}$. This splitting term is nonzero only when both $h_{eff}$ and $\delta t$ are nonzero. This result shows that both the effects of nonmagnetic ions, which break translational symmetry between A/B sublattices, and the antiparallel exchange field between A- and B-type magnetic ions are essential for the emergence of spin splitting in altermagnetism, distinguishing it from the antiferromagnetic state. This spin splitting is typically on the order of an electron volt (eV), indicating strong spin-dependent anisotropy in the band structure within reciprocal momentum space. Such pronounced spin-dependent anisotropy in altermagnetism has potential applications in spintronics and spin-caloritronics [5].

The minimal model further confirms that spin and lattice symmetries determine the crystal momenta at which spin degeneracy occurs. Equations (1) and (3) indicate that spin degeneracy should be present at the zone boundary, zone center, and along the $k_{x/y} = 0$ lines. The first two instances of spin degeneracy arise from the translational symmetry of the magnetic unit cell. In contrast, the spin degeneracy along the $k_{x/y} = 0$ lines is governed by a symmetry operator that combines the following operations: a global spin flip at all lattice sites, a translation between nearest-neighbor A and B sublattices, and a fourfold rotational symmetry associated with the point group symmetry of the magnetic ions. This symmetry-driven spin degeneracy, along with potential accidental spin degeneracy arising from orbital and lattice degrees of freedom, should contribute to the anomalous Hall coefficient and topological currents [6,17,18].

# V. Resulting band structure

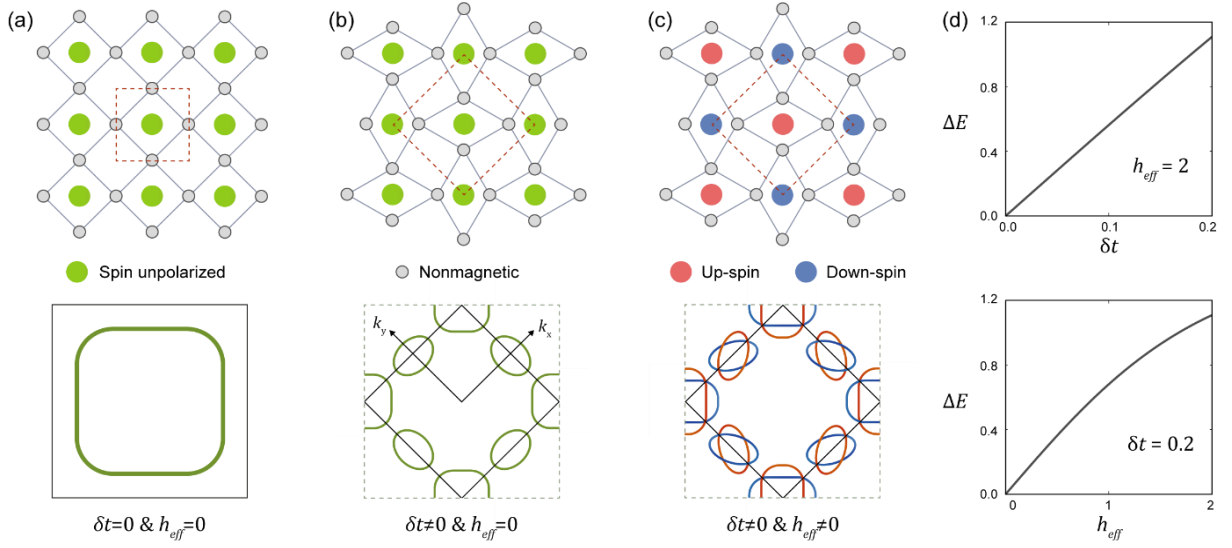

**Figure 4** Fermi surface symmetry derived from the minimal model. (a) Non-magnetic state without distortion of non-magnetic ion cages. (b) Non-magnetic state with distortion of non-magnetic ion cages. (c) AM state. Upper panels show the crystal structure while lower panels are corresponding schematic Fermi surfaces. The red dashed square is the unit cell of the crystal structure. The black solid square in the lower panels is the first Brillouin zone. For (a) and (b), green curves show spin un-polarized Fermi surfaces while the red and blue curves are for spin-up and spin-down Fermi surfaces, respectively. Corresponding $\delta t$ and $h_{eff}$ are noted at the bottom. (d) Spin splitting $\Delta E$ at $(k_x, k_y) = \left(\frac{\pi}{2}, \frac{\pi}{2}\right)$ as a function of $\delta t$ (upper, with $h_{eff} = 2$) and $h_{eff}$ (lower, with $\delta t = 0.2$) from Eqn (3). The unit of energy is $t_1$ in Eqn (2).

With the minimal model, we can obtain the band structure and investigate the effects of $\delta t$ and $h_{eff}$. Figure 4 presents crystal structure along with schematic Fermi surfaces within the minimal model for various states. In the case of a non-spin-polarized state without distortion of the non-magnetic ions, as shown in Fig. 4(a), the translational symmetry between the nearest-neighbor magnetic ions is preserved, resulting in the absence of spin splitting. Consequently, a single Fermi surface emerges within the enlarged Brillouin zone. This scenario corresponds to the conditions $\delta t = 0$ and $h_{eff} = 0$ in Eqn (1) of the minimal model. For a non-spin-polarized state with distortion of the non-magnetic ions, as depicted in Fig. 4(b), the translational symmetry between nearest-neighbor magnetic ions is broken, while spin splitting remains absent. As a result, two Fermi surfaces emerge in the folded Brillouin zone, but without spin splitting. This configuration corresponds to $\delta t \neq 0$ and $h_{eff} = 0$ in Eqn (1) of the minimal model.

In the case of an altermagnetic state, as illustrated in Fig. 4(c), the conditions $\delta t \neq 0$ and $h_{eff} \neq 0$ in Eqn (1) are realized. Consequently, four Fermi surfaces with spin splitting emerge within the folded Brillouin zone. As Eqns (1) and (3) show, nodal points for spin degeneracy appear at the zone boundary and along the $k_{x/y} = 0$ lines. The presence of these four Fermi surfaces with nodal points suggests that both the translational symmetry breaking induced by distorted non-magnetic ions and the alternating spin polarization of magnetic ions are essential conditions for the emergence of altermagnetism. In Fig. 4(d), it is confirmed that both terms, $\delta t$ and $h_{eff}$, promote the spin splitting in the altermagnetic state and that both should be non-zero for the emergence of the spin splitting.

# VI. A case study: Altermagnetic wave function of MnF$_2$

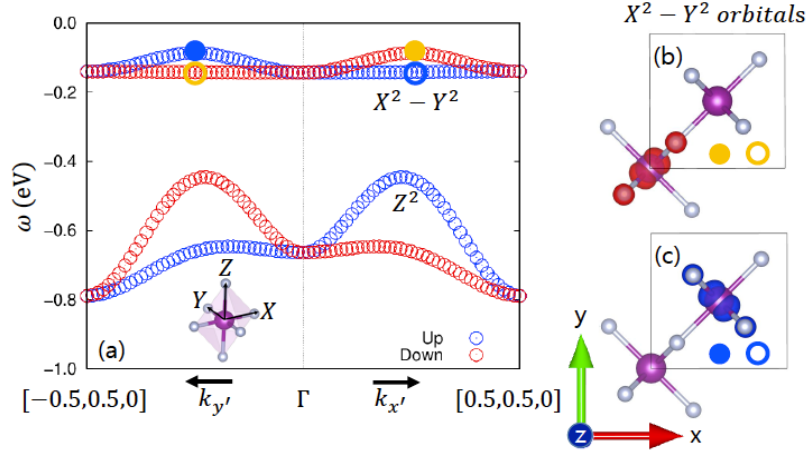

**Figure 5** Electronic structures of altermagnetic MnF$_2$ within the DFT-GGA framework. (a) Band structure of altermagnetic MnF$_2$. The topmost and second topmost valence bands are from the $X^2 - Y^2/Z^2$ orbitals of Mn($d$) in the distorted octahedra, where $Z$, $X$, and Y are local coordinates with $Z$ ($X,Y$) fixed along the elongated (shortened) Mn-F bond direction of the octahedron (see the inset for the definition). Blue and red colors present spin up and down, respectively. (b) and (c) Bloch wavefunctions' contribution to the charge density at a specific crystal momentum of a band as marked by orange and blue circles, respectively, in panel (a).

One may investigate wave functions of real material to gain further insights on the effect of the anisotropic hopping in AM. Figure 5 illustrates the electronic structure of a representative altermagnetic material MnF$_2$, calculated within the framework of density functional theory using the generalized gradient approximation for the exchange-correlation potential (DFT-GGA). In Fig. 5(a), we plot the electronic structure along the $k_{x'}$ and $k_{y'}$ directions for the topmost and second topmost bands, which originate from the Mn $e_g$ orbitals in an octahedral coordination of $X$ and $Y$. The characteristic spin splitting with a zero net magnetic moment, a hallmark of altermagnetism, is clearly evident.

Figures 5(b) and (c) depict the wave functions' contribution to the charge density for the topmost bands (derived from the $X^2 - Y^2$ orbitals) at the crystal momenta [−0.25,0.25,0] and [0.25,0.25,0]. These plots provide clear evidence of the role of non-magnetic ions in the emergence of altermagnetism. The $X^2 - Y^2$ wave functions conform to the point group symmetries of each sublattice, while the Mn($d$)-F($p$) hybridization follows these symmetries, breaking the translational symmetry between the sublattices. As illustrated in Fig. 3, this broken translational symmetry gives rise to the $\delta t$ term in Eqn (2). For instance, for the $X^2 - Y^2$ orbital driven band, the spin up (blue) wave function is elongated along [-0.5,0.5,0] while the spin down (red) wave function is elongated along [0.5,0.5,0], corresponding to the more dispersive direction for both cases. The situation is similar for the $Z^2$ orbital driven bands except that the dispersive direction is opposite to that of the $X^2 - Y^2$ orbital case. Furthermore, the Mn-F bond length is elongated along the $Z$ direction, which gives rise to a larger spin splitting for the $Z^2$ orbital driven band in comparison to the $X^2 - Y^2$ orbital driven band.

Furthermore, due to the strong magnetic exchange energy ($h_{eff}$) associated with the Mn($d^5$) configuration, Mn($d$)-F($p$) hybridization induces spin splitting in the electronic structure. The wave function analysis in Figures 5(b) and (e) confirms that the system preserves the $[C_2||C_{4z}\tau_{AB}]$ symmetry, which is dictated by the interplay between non-magnetic ions and magnetic ordering. This symmetry ensures a zero net magnetic moment while giving rise to a $d$-wave nodal spin-resolved electronic structure.

## Acknowledgements

This work was supported by the National Research Foundation of Korea (NRF) grant funded by the Korean government (MSIT) (No. 2022R1A3B1077234) and Global Research Development Center Cooperative Hub Program (GRDC) through NRF (RS-2023-00258359). This work was also supported by the Institute of Applied Physics, Seoul National University. MK was supported by Korea Institute for Advanced Study (KIAS) individual Grants (No. CG083502). The DFT calculation is supported by the Center for Advanced Computation at KIAS.